\documentclass[preprint,proceedings]{rmaa}
\usepackage{psfig}
\usepackage{epsf}

\usepackage{paralist}

\usepackage{psfrag,color}

\newcommand{\cmm}{\,{\rm cm}^{-2}}

\newcommand{\Lya}{Ly$\alpha\ $}

\makeatletter

\newenvironment{figurehere}
  {\def\@captype{figure}}
  {}
\makeatother

\def\kms{\,{\rm km\,s^{-1}}}
\def\kmsmpc{\,{\rm km\,s^{-1}\,Mpc^{-1}}}
\def\msun{\,{\rm M_\odot}}

\def\spose#1{\hbox to 0pt{#1\hss}}
\def\lta{\mathrel{\spose{\lower 3pt\hbox{$\mathchar"218$}} \raise 2.0pt\hbox{$\mathchar"13C$}}}
\def\gta{\mathrel{\spose{\lower 3pt\hbox{$\mathchar"218$}} \raise 2.0pt\hbox{$\mathchar"13E$}}}
\def\cmm{\,{\rm cm}^{-2}}

\def\lya{Ly$\alpha\ $}
\def\uvunits{{\rm\,ergs\,cm^{-2}\,s^{-1}\,Hz^{-1}\,sr^{-1}}}

 \def\ni{\noindent}
\def\HI{\hbox{H~$\scriptstyle\rm I\ $}}
\def\HII{\hbox{H~$\scriptstyle\rm II\ $}}

\def\CIV{\hbox{C~$\scriptstyle\rm IV\ $}}
\def\SiIV{\hbox{Si~$\scriptstyle\rm IV\ $}}

\def\nHI{{\rm HI}}
\def\nH{{\rm H}}

\def\ni{{\noindent}}

\def\vir{{\rm vir}}
\def\igm{{\rm IGM}}

\SetYear{2002}
\SetConfTitle{Galaxy Evolution: Theory and Observations}

\title{The Era of Reionization} 

\author{Piero Madau
\altaffil{Department of Astronomy and Astrophysics, University of California,
Santa Cruz, CA 95064, USA.}}

\shortauthor{Madau}
\shorttitle{Reionization}

\fulladdresses{
\item Piero Madau: University of California, 1156 High Street, Santa 
Cruz, CA 95064, USA (pmadau@ucolick.org).}
\listofauthors{P. Madau}
\indexauthor{Madau, P.}

\abstract{
In popular cold dark matter cosmological scenarios, stars may have 
first appeared in significant numbers around a redshift of 10 or so, 
as the gas within protogalactic halos with virial temperatures $T_\vir\gta 
10^{4.3}\,$K (corresponding to masses comparable to those of present--day 
dwarf ellipticals) cooled rapidly due to atomic processes and fragmented.
It is this `second generation' of subgalactic stellar systems, aided 
perhaps 
by an early population of accreting black holes in their nuclei, which  
may have generated the 
ultraviolet radiation and mechanical energy that ended the cosmic ``dark 
ages''
and reheated and reionized most of the hydrogen in the universe by a 
redshift of 6. The detailed history of the universe during and soon after 
these crucial formative stages depends on the power--spectrum of density
fluctuations on small scales and on a complex network of poorly understood
`feedback' mechanisms, and is one of the missing link in galaxy formation 
and evolution studies. The astrophysics of the epoch of first light is 
recorded in the thermal state, ionization degree, and chemical composition 
of the intergalactic medium, the main repository of baryons at high 
redshifts.}

\addkeyword{cosmology}
\addkeyword{diffuse radiation}
\addkeyword{intergalactic medium}

\begin{document}
% Typeset article header
\maketitle

\section{Introduction}

At epochs corresponding to $z\sim 1000$ the intergalactic medium (IGM)
is expected to recombine and remain neutral until sources of radiation and
heat develop that are capable of
reionizing it. The detection of transmitted flux shortward of the \lya
wavelength in the spectra of sources at $z\gta 5$ implies that the hydrogen
component of this IGM was ionized at even higher redshifts. 
The increasing thickening of the \Lya forest recently measured in the 
spectra of SDSS $z\sim 6$ quasars (Becker \etal 2001; Djorgovski \etal 
2001) may be the signature of the trailing edge of the cosmic reionization 
epoch. 
It is clear that substantial sources of ultraviolet photons and mechanical
energy, like young star--forming galaxies, were already present back then. 
The reionization of intergalactic hydrogen at $z\gta 6$ is unlikely to have 
been accomplished by quasi--stellar sources: the observed dearth 
of luminous 
optical and radio--selected QSOs at $z>3$ (Shaver \etal 1996; Fan \etal 2001), 
together with the detection of substantial Lyman--continuum flux in a composite 
spectrum of Lyman--break galaxies at $\langle z\rangle=3.4$ (Steidel 
et al. 2001), both may support the idea that massive stars in galactic and 
subgalactic systems -- rather than quasars -- reionized the hydrogen component 
of the IGM when the universe was less than 5\% of its current age, and 
dominate the 1 ryd metagalactic flux at all redshifts greater than 3.
An episode of pregalactic star formation may also provide a possible 
explanation
for the widespread existence of heavy elements (like carbon, oxygen, and
silicon) in the IGM. There is mounting evidence that the 
double reionization of helium may have occurred later, at a redshift of 3 or 
so (see Kriss et al. 2001, and references therein): this is likely due to the
integrated radiation emitted above 4 ryd by QSOs.

Establishing what ended the dark ages and when is important 
for determining the impact of cosmological reionization and reheating 
on several key cosmological issues, from
the role reionization plays in allowing protogalactic objects to cool and
make stars, to determining the thermal state of baryons at high redshifts
and the small--scale structure in the temperature
fluctuations of the cosmic microwave background. Conversely, probing the
reionization epoch may provide a means for constraining competing models for
the formation of cosmic structures: for example, popular modifications of 
the CDM paradigm that attempt to improve over CDM by suppressing the primordial 
power--spectrum on small scales, like warm dark matter (WDM), are known to 
reduce the number of collapsed halos at high redshifts and make it 
more difficult to reionize the universe (Barkana \etal 2001).       
In this talk I will summarize some recent theoretical developments in 
understanding the astrophysics of the epoch of first light and the impact 
that some of the earliest generations of stars, galaxies, and 
black holes in the universe may have had on the IGM.

\section{Reionization by massive stars}

In a cold dark matter (CDM) universe, structure formation is a hierarchical process in
which non linear, massive structures grow via the merger of smaller initial
units. Large numbers of low--mass galaxy halos are expected to form at
early times in these popular theories, leading to an era of reionization, reheating, and 
chemical enrichment.  Most models predict that intergalactic hydrogen
was reionized by an early generation of stars or accreting black holes at 
$z=7-15$. One should note, however, that while numerical N--body$+$hydrodynamical simulations
have convincingly shown that the IGM does fragment into structures at
early times in CDM cosmogonies (e.g. Cen \etal 1994;
Zhang \etal 1995; Hernquist \etal 1996),
the same simulations are much less able to predict the efficiency with
which the first gravitationally collapsed objects lit up the universe
at the end of the dark age. 

The scenario that has received the most theoretical studies is one 
where hydrogen is photoionized by the UV radiation emitted either by quasars or
by  stars with masses $\gta 10\,\msun$ (e.g. Shapiro \& Giroux 1987; Haiman 
\& Loeb 1998; Madau \etal 1999; Chiu \& Ostriker 2000; Ciardi \etal 
2000), rather than ionized by collisions with 
electrons heated up by, e.g. supernova--driven winds from early pregalactic
objects.  In the former case a high degree of ionization requires
about $13.6\times (1+t/\bar t_{\rm rec})\,$eV per hydrogen atom, where 
$\bar t_{\rm rec}$ is the volume--averaged hydrogen recombination timescale, 
$t/\bar t_{\rm rec}$ being much greater than unity already at 
$z\sim 10$ according to numerical simulations.
Collisional ionization to a neutral fraction of only a  few parts in $10^{5}$ requires a 
comparable energy input, i. e. an IGM temperature close to $10^5\,$K or about $25\,$eV per atom.

Massive stars will deposit both radiative and mechanical energy into the 
interstellar medium of protogalaxies. A complex network of `feedback' 
mechanisms is likely at work in these systems, as the gas in shallow potential
is more easily blown away (Dekel \& Silk 1986; Tegmark \etal 1993;
Mac Low \& Ferrara 1999; Mori et al. 2002) thereby quenching star formation. 
Furthermore, as the 
blastwaves produced by supernova explosions reheat the surrounding 
intergalactic gas and enrich it with newly formed heavy elements (see below),
they can inhibit the formation of surrounding low--mass galaxies 
due to `baryonic stripping' (Scannapieco \etal 2002). It is therefore 
difficult to establish whether an early input of mechanical energy will 
actually play a major role in determining the thermal and ionization state 
of the IGM on large scales. What can be easily shown is that, during 
the evolution of a a `typical' 
stellar population, more energy is lost in ultraviolet radiation than in
mechanical form. This is because in nuclear burning from zero to solar 
metallicity ($Z_\odot=0.02$), the energy radiated per 
baryon is $0.02\times 0.007\times m_\nH c^2$, with about one third of it 
going into 
H--ionizing photons. The same massive stars that dominate the UV light 
also explode as supernovae (SNe), returning most of the metals to the 
interstellar medium and 
injecting about $10^{51}\,$ergs per event in kinetic energy. For a Salpeter 
initial mass function (IMF), one has about one SN every $150\,\msun$ of baryons
that forms stars. The mass fraction in mechanical energy is then approximately
$4\times 10^{-6}$, ten times lower than the fraction released in photons
above 1 ryd. 

The relative importance of photoionization versus shock ionization will 
depend, however, on the efficiency with which radiation and mechanical 
energy actually escape into the IGM.    
Consider, for example, the case of an early generation of subgalactic 
systems collapsing
at redshift 9 from 2--$\sigma$ fluctuations. At these epochs their dark matter  
halos would have virial radii $r_{\rm vir}=0.75 h^{-1}$ 
Kpc and circular velocities $V_c(r_{\rm vir})=25\,\kms$,
corresponding in top--hat spherical collapse to a virial 
temperature $T_\vir=0.5\mu m_p V_c^2/k\approx 10^{4.3}\,$K
and halo mass $M=0.1V_c^3/GH\approx 10^8 h^{-1}\, \msun$.\footnote{This 
assumes an Einstein--de Sitter universe with Hubble constant 
$H_0=100\,h\,\kmsmpc$.}~ Halos in this mass
range are characterized by very short dynamical timescales (and even shorter
gas cooling times due to atomic hydrogen) and may therefore form stars in 
a rapid but intense burst before SN `feedback' quenches further star formation.
For a star formation efficiency of $f=0.1$, $\Omega_bh^2=0.02$\footnote{Here $\Omega_b$ is 
the baryon density parameter, and $f\Omega_b$ is the 
fraction of halo mass converted into stars.}, $h=0.5$, and a Salpeter IMF,
the explosive output of $10,000$ SNe will inject an energy $E_0\approx 
10^{55}\,$ergs. This is roughly a hundred times higher than the gas binding
energy: a significant fraction of the halo gas will then be lifted out of the 
potential well (`blow--away') and shock the intergalactic medium
(Madau \etal 2001).
If the explosion occurs at cosmic time $t=4\times 10^8\,$yr (corresponding in 
the adopted cosmology to $z=9$), at time $\Delta t=0.4t$ after the event 
it is a good 
approximation to treat the cosmological blast wave as adiabatic, with  
proper radius given by the standard Sedov--Taylor self--similar solution,
$$ 
R_s\approx \left(\frac{12\pi G \eta E_0}{\Omega_b}\right)^{1/5}t^{2/5}\Delta 
t^{2/5} \approx 17\,{\rm kpc}.  \eqno(1)  
$$
Here $\eta\approx 0.3$ is the fraction of the available SN energy that is 
converted into kinetic energy of the blown--away material (Mori \etal 2002).
At this instant the shock velocity relative to the Hubble flow is  
$$
v_s\approx 2R_s/5\Delta t\approx 40\,\kms,  \eqno(2)  
$$
lower than the escape velocity from the halo center. The gas temperature 
just behind the shock front is $T_s=3\mu m_pv_s^2/16k 
\gta 10^{4.3}\,$K, enough to ionize all 
incoming intergalactic hydrogen. At these redshifts, it is the onset of 
Compton cooling off 
cosmic microwave background photons that ends the adiabatic stage of blast 
wave propagation; the shell then enters the snowplough phase and is finally 
confined by the IGM pressure. 
According to the Press--Schechter formalism, the 
comoving abundance of collapsed dark halos with mass $M=10^8\,h^{-1}\,M_\odot$ 
at $z=9$ is $dn/d\ln M\sim 80\,h^3\,$Mpc$^{-3}$, corresponding to a mean 
proper distance between neighboring halos of $\sim 15\,h^{-1}\,$kpc. With the 
assumed star formation efficiency, only a small fraction, about 4 
percent, of the total stellar mass inferred today (Fukugita
\etal 1998) would actually form at these early epochs. 
Still, our simple analysis shows that the blast waves from such 
a population of pregalactic objects could 
drive vast portions of the IGM to a significantly higher 
adiabat, $T_\igm \sim 10^5\,$K, than expected from photoionization, so as 
to `choke off' the collapse of further $M\lta 10^9\,h^{-1}\msun$ systems by 
raising the cosmological Jeans mass. In this sense the process may 
be self--regulating. 

The thermal history of expanding intergalactic primordial gas at the mean 
density is plotted in Figure 1 as a function of redshift for a number 
of illustrative cases. The code we have used includes the relevant cooling and
heating processes and follows the non--equilibrium evolution of hydrogen and
helium ionic species in a cosmological context. 

\begin{figurehere}
\vspace{-0.6cm}
\centerline{
\psfig{figure=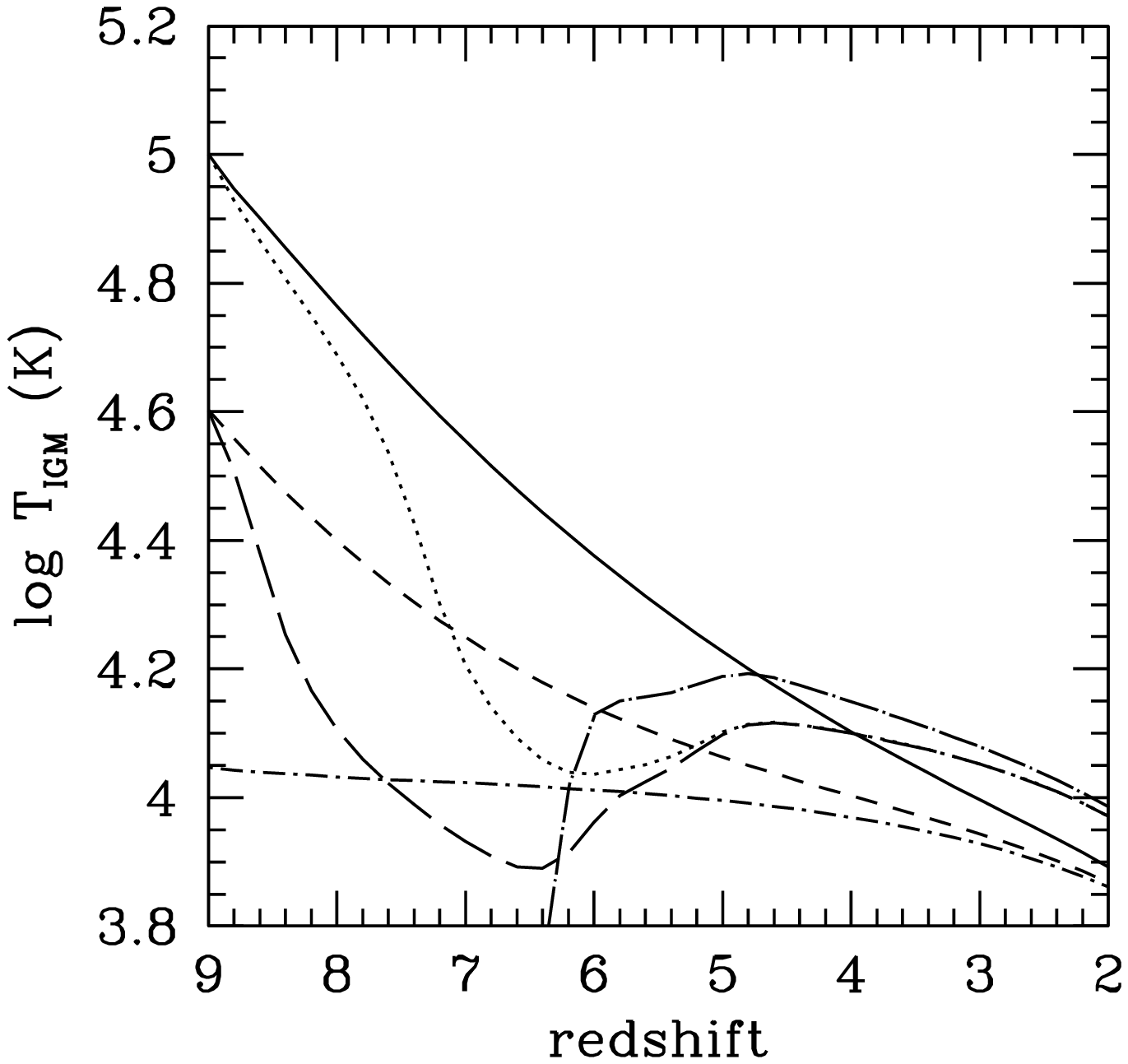,width=2.7in}}
\vspace{-0.9cm}
\caption{\footnotesize Thermal history of intergalactic gas at the mean 
density in an Einstein--de Sitter universe with $\Omega_bh^2=0.02$ and $h=0.5$.
{\it Short dash--dotted line:} temperature evolution when the
only heating source is a constant ultraviolet (CUV) background of intensity
$10^{-22}\,\uvunits$ at 1 Ryd and power--law spectrum with energy 
slope $\alpha=1$.
{\it Long dash--dotted line:} same for the time--dependent quasar ionizing
background as computed by Haardt \& Madau (1996; HM).
{\it Short--dashed line:} heating due to a CUV background  but with an initial
temperature of $4\times 10^4\,$K at $z=9$ as expected from an early era
of pregalactic outflows.
{\it Long--dashed line:} same but for a HM background. {\it Solid line:} heating
due to a CUV background  but with an initial temperature of $10^5\,$K
at $z=9$.
{\it Dotted line:} same but for a HM background.
}
\vspace{0.2cm}
\end{figurehere}

\ni The gas is allowed to interact
with the CMB through Compton cooling and either with a time--dependent QSO ionizing
background as computed by Haardt \& Madau (1996) or with a
time--independent metagalactic flux of intensity $10^{-22}\,\uvunits$ at 1 Ryd
(and power--law spectrum with energy slope $\alpha=1$). The temperature of the
medium at $z=9$ -- where we start our integration -- has been either
computed self--consistently from photoheating or
fixed to be in the range $10^{4.6}-10^5\,$K expected from SN--driven
bubbles with significant filling factors (see below). The various curves show that the
temperature of the IGM at $z=3-4$ will retain little memory of
an early era of pregalactic outflows and preheating, and be consistent
with that expected from photoionization.

\section{Feedback and pregalactic enrichment}

Understanding the origin of the chemical elements, following the increase
in their abundances with cosmic time, and uncovering the processes
responsible for distributing the products of stellar nucleosynthesis
over very large distances are all key aspects of the evolution of gaseous 
matter in the universe. One of the major discoveries with {\it Keck} concerning the IGM has 
been the identification of metal absorption lines associated with many of the 
\Lya forest systems. The detection of 
measurable \CIV and \SiIV absorption lines in clouds with \HI column densities
as low as $10^{14.5}\,\cmm$ implies a minimum universal metallicity relative
to solar in the range $[-3.2]$ to $[-2.5]$ at $z\sim 3$ (Songaila 1997).
There is no indication in the data of a turnover in the \CIV column density
distribution down to $N_{\rm CIV}\lta 10^{11.7}\, \cmm$ ($N_\nHI\lta
10^{14.2}\, \cmm$, Ellison \etal 2000).

From a theoretical perspective, it is unclear whether the existence of heavy 
elements in the IGM at $z=3-3.5$ points to an early ($z>6$) enrichment epoch by 
low--mass subgalactic systems (Madau \etal 2001), or is rather due to 
late pollution by the population of star--forming galaxies known to be already in 
place at $z=3$. The Press--Schechter theory for the evolving mass function of dark 
matter halos predicts a power--law dependence, $dN/d\ln m\propto m^{(n_{\rm eff}-3)/6}$, 
where $n_{\rm eff}$ is the effective slope of the CDM power spectrum, $n_{\rm eff}\approx 
-2.5$ on subgalactic scales. As hot, metal--enriched gas from SN--driven winds 
escapes its host halo, shocks the IGM, and eventually forms a blast wave, it sweeps a
region of intergalactic space the volume of which increases with the $3/5$ power of the
injected energy $E_0$ (in the adiabatic Sedov--Taylor phase).
The total fractional volume or porosity, $Q$, filled by these `metal bubbles'
per unit explosive energy density $E_0\,dN/d\ln m$ is then
$ Q\propto E_0^{3/5}\,dN/d\ln m\propto (dN/d\ln m)^{2/5}\propto m^{-11/30}$.
Within this simple scenario it is the star--forming objects with the smallest
masses which will arguably be the most efficient pollutant of the IGM on
large scales. Metal--enriched material from SN ejecta may be far more easily
accelerated to escape velocities in the shallow 
potential wells of subgalactic systems at $z>6$.
Late enrichment may also encounter problems in explaining the kinematic quiescence 
of \CIV lines; the observed small scale properties of the IGM at $z\sim 3$
appear consistent with \CIV absorbers being the result of ancient pregalactic 
outflows (Rauch \etal 2001). 
According to numerical hydrodynamics simulations of structure formation in the 
IGM, the metals associated with $\log N_\nHI \lta 14$ filaments fill a fraction $\gta 5\%$ 
of intergalactic space, and are
therefore far away from the high overdensity peaks where galaxies form, gas
cools, and
star formation takes place. Their chemical enrichment may then reflect more
uniform (i.e. `early') rather than in--situ (i.e. `late') metal pollution.
The case for pregalactic enrichment may have been recently strenghtened by the
observation of an invariant \CIV column density distribution 
throughout the redshift range $2<z<5$ (Songaila 2001).  

\bigskip
\begin{figurehere}
\centerline{\psfig{figure=pmadau_fig2.ps,height=11cm}} 
\vspace{0.5cm}
\caption{\footnotesize 
Snapshots of the logarithmic number density of the gas
at five different elapsed times for our Case 1 simulation run. The three
panels in each row show the spatial density distribution in the $X-Y$
plane on the nested grids.
The three columns in each figure depict the time evolution from about 6 Myr to
up to 180 Myr. Along a given row, the leftmost panel refers to grid L5 (linear
size 48 kpc), the central one to grid L3 (linear size 12 kpc), and the rightmost
panel refers to the grid L1 (linear size 3 kpc). The density range is
$-5 \le \log~(n/{\rm cm}^{-3}) \le 1$.
The halo gas is assumed to be initially in hydrostatic equilibrium and 
non self--gravitating. (From Mori \etal 2002.)}
\vspace{0.5cm}
\end{figurehere} 

\ni To simulate the process of blow--away we have developed a 3D Eulerian code that
solves the hydrodynamic equations for a perfect fluid in Cartesian geometry.
To deal with very different length scales in our simulation we have 
adopted a `nested grid method' with six levels of 
fixed Cartesian grids. The grids are connected by the transfer of 
conserved variables, and are centered within each other, with the finest 
covering the whole galaxy halo.
Since the cell number is the same ($128\times 128\times 128$) for
every level L$n$ ($n=1, 2, ..., 6$), the minimum resolved scale is
about 22 pc and the size of the coarsest grid is 96 kpc. Thus, the scheme has
a wide dynamic range in the space dimension.

The results of a numerical simulation (run on massive parallel supercomputers at the
Center for Computational Physics, Tsukuba University) are depicted in Figure 2. 
In this run about 10,000 SNe explode in a high redshift protohalo 
(corresponding to a star formation efficiency of about 10\% for a Salpeter IMF). 
Our algorithm for simulating SN feedback improves upon previous treatments
in several ways. OB associations are distributed as a function of gas
density according to a Schmidt--type law ($\propto \rho^\alpha$) using a 
Monte--Carlo procedure. After a main sequence lifetime, all stars more massive
than 8 M$_{\odot}$ explode instantaneously injecting an energy of $10^{51}$
ergs, and their outer layers are blown out leaving a compact remnant of 1.4
M$_{\odot}$. Therefore SNe inject energy (assumed in pure thermal form) and mass into
the interstellar medium: these are supplied to a sphere 
corresponding to the radius of a SN remnant in a uniform ambient medium of density
$n$ and in the adiabatic Sedov--Taylor phase; the expansion velocity
is also self--consistently calculated from such solution. The gas then starts 
cooling immediately according to the adopted gas cooling function.
While the time--step $\Delta t$ is controlled by the Courant condition, if there 
are more than two SN events in a OB association we decrease $\Delta t$ until 
only one explosion per association occurs during the time--step. 

Snapshots of the gas density distribution in a 
planar slice of the nested grids are shown for an extended stellar distribution case
($\alpha=1$). After a few Myr from the beginning of the simulation the most 
massive stars explode as SNe and produce expanding hot bubbles surrounded
by a cooling dense ($n \approx 1$~cm$^{-3}$) shell.
As the evolution continues, a coherent and increasingly spherical shell
expanding into the IGM is eventually formed. The shell contains
a large fraction of the halo gas that has been swept--up during the evolution.
The final bottom row of the simulation figure shows the final stages of
the evolution: the shell is now nearly spherically symmetric, its interior being
filled with warm ($T\lta 10^6$~K)
gas at a very low density $n\lta 10^{-4}$~cm$^{-3}$, i.e. slightly below the
mean value for the IGM. At the end of the simulations, the shell is
still sweeping  out IGM material; its radius and  velocity are 21 kpc and 
26 km sec$^{-1}$ at $t=250\,$Myr. 
Using momentum conservation one can estimate the final radius of the shell
to be close to 25 kpc. This is about 17 times the virial radius of the halo!

It is clear then that SN--driven pregalactic outflows may be an efficient mechanism
for spreading metals around. The collective explosive output
of about ten thousands SNe per $M\gta 10^8\,h^{-1}\,\msun$ halo at these early
epochs could then pollute the entire intergalactic space to a mean metallicity
$\langle Z\rangle=\Omega_Z/\Omega_b\gta 0.003$ (comparable to the levels
observed in the \lya forest at $z\approx 3$) without much perturbing the IGM
hydrodynamically, i.e. producing large variations of the baryons relative
to the dark matter. The significance of these results goes well beyond early metal
enrichment. This is because, since the cooling time of collisionally ionized
high density gas in small halos at high redshifts is much shorter than the
then Hubble time, virtually
all baryons are predicted to sink to the centers of these halos in the absence
of any countervailing effect (White \& Rees 1978). Efficient feedback is then
necessary in hierarchical clustering scenarios to avoid this `cooling
catastrophe',
i.e. to prevent too many baryons from turning into stars as soon as the first
levels of the hierarchy collapse. The required reduction of the stellar
birthrate in halos with low circular velocities may naturally result from
the heating and expulsion of material due to OB stellar winds and repeated
SN explosions from the first burst of star formation. 

\section{Probing the epoch of first light at 21--cm}

\ni  Prior to the epoch of full reionization, the intergalactic medium and 
gravitationally collapsed systems may be detectable in emission or absorption 
against the CMB at the frequency corresponding to the redshifted 21--cm line 
(associated with the spin--flip transition from the triplet to the singlet 
state of neutral hydrogen.) In general, 21--cm spectral features
are expected to display angular structure as well as structure in redshift 
space due to inhomogeneities in the gas density field, hydrogen ionized 
fraction, and spin temperature.  Several different signatures have been 
investigated in the recent literature: (a) the fluctuations in the redshifted
21--cm emission induced by the gas density inhomogeneities that develop
at early times in CDM-dominated cosmologies (Madau \etal 1997; Tozzi \etal 
2000) and by virialized ``minihalos''  with $T_\vir<10^4\,$K (Iliev \etal
2002);
(b) the sharp absorption feature in the radio sky due to 
the rapid rise of the \Lya continuum background that marks the birth of the 
first UV sources in the universe (Shaver \etal 1999); (c) the 21--cm narrow 
lines generated in absorption against very high--redshift radio sources 
by the neutral IGM (Carilli \etal 2002) and by intervening minihalos  
and protogalactic disks (Furlanetto \& Loeb 2002).

A quick summary of the physics of 21--cm radiation will illustrate 
the basic ideas and unresolved issues behind these studies.
The emission or absorption of 21--cm photons from a neutral IGM is
governed by the hydrogen spin temperature $T_S$ defined by $n_1/n_0=3
\exp(-T_*/T_S)$,
where $n_0$ and $n_1$ are the singlet and triplet $n=1$ hyperfine levels
and $k_BT_*=5.9\times 10^{-6}\,$eV is the energy of the 21--cm transition. 
In the presence of only the CMB radiation with
$T_{\rm CMB}=2.73\,(1+z)\,$K, the spin states will reach thermal equilibrium
with the CMB on a timescale of $T_*/(T_{\rm CMB}A_{10})\approx 3\times10^5\,
(1+z)^{-1}\,$yr ($A_{10}=2.9\times 10^{-15}\,$s$^{-1}$
is the spontaneous decay rate of the hyperfine transition of atomic hydrogen),
and intergalactic \HI will produce neither an absorption nor an emission
signature. A mechanism is required that decouples $T_S$ and $T_{\rm CMB}$,
e.g. by coupling the spin temperature instead to the kinetic temperature
$T_K$ of the gas itself. Two mechanisms are
available, collisions between hydrogen atoms (Purcell \& Field 1956)
and scattering by \Lya photons (Field 1958). The
collision--induced coupling between the spin and kinetic temperatures
is dominated by the spin--exchange process between the colliding
hydrogen atoms. The rate, however, is too small for realistic IGM
densities at the redshifts of interest, although collisions may be
important in dense regions with $\delta\rho/\rho\gta 30[(1+z)/10]^{-2}$, like
virialized minihalos.

In the low density IGM, the dominant mechanism is the scattering of continuum
UV photons redshifted by the Hubble expansion into local \Lya photons. The
many scatterings mix the hyperfine levels of neutral hydrogen in its
ground state via intermediate transitions to the $2p$ state, the
Wouthuysen--Field process. As the neutral IGM is highly opaque to resonant 
scattering, the shape of the continuum radiation spectrum
near \Lya will follow a Boltzmann distribution with a temperature given by
the kinetic temperature of the IGM (Field 1959). In this case the spin
temperature of neutral hydrogen is a weighted mean between the matter and
CMB temperatures. 
There exists then a critical value of the background flux of \Lya photons
which, if greatly exceeded, would drive the spin temperature away from 
$T_{\rm CMB}$. 

While the microphysics is well understood, our understanding of the 
astrophysics of 21--cm tomography is still poor.  
In Tozzi \etal (2000) we used N--body cosmological simulations and, 
assuming a fully neutral medium with $T_S\gg T_{\rm CMB}$, showed that prior to 
reionization the same network of sheets and filaments (the `cosmic web') that 
gives rise to the \lya forest at $z\sim 3$ should lead to fluctuations in the 
21--cm brightness temperature at higher redshifts (Figure 3). 
At 150 MHz ($z=8.5$), for
observations with a bandwidth of 1 MHz, the root mean square fluctuations
should be $\sim 10\,$ mK at $1'$, decreasing with scale.
Because of the smoothness of the CMB sky, fluctuations in the 21--cm
radiation will dominate the CMB fluctuations by about 2 orders of magnitude
on arcmin scales. The search at 21--cm for the epoch of first light has 
become one of the main science drivers of the {\it LOw Frequency ARray} 
({\it LOFAR}; see http://www.astron.nl/lofar/science/).  While remaining 
an extremely challenging project due to foreground contamination from 
extragalactic radio sources  (Di Matteo \etal 2002), the detection and 
imaging of these small--scale structures with {\it LOFAR} is a tantalizing 
possibility within range of the thermal noise of the array. 

\begin{figurehere}
\vspace{0.7cm}
\hspace{0.cm}
\centerline{
\psfig{figure=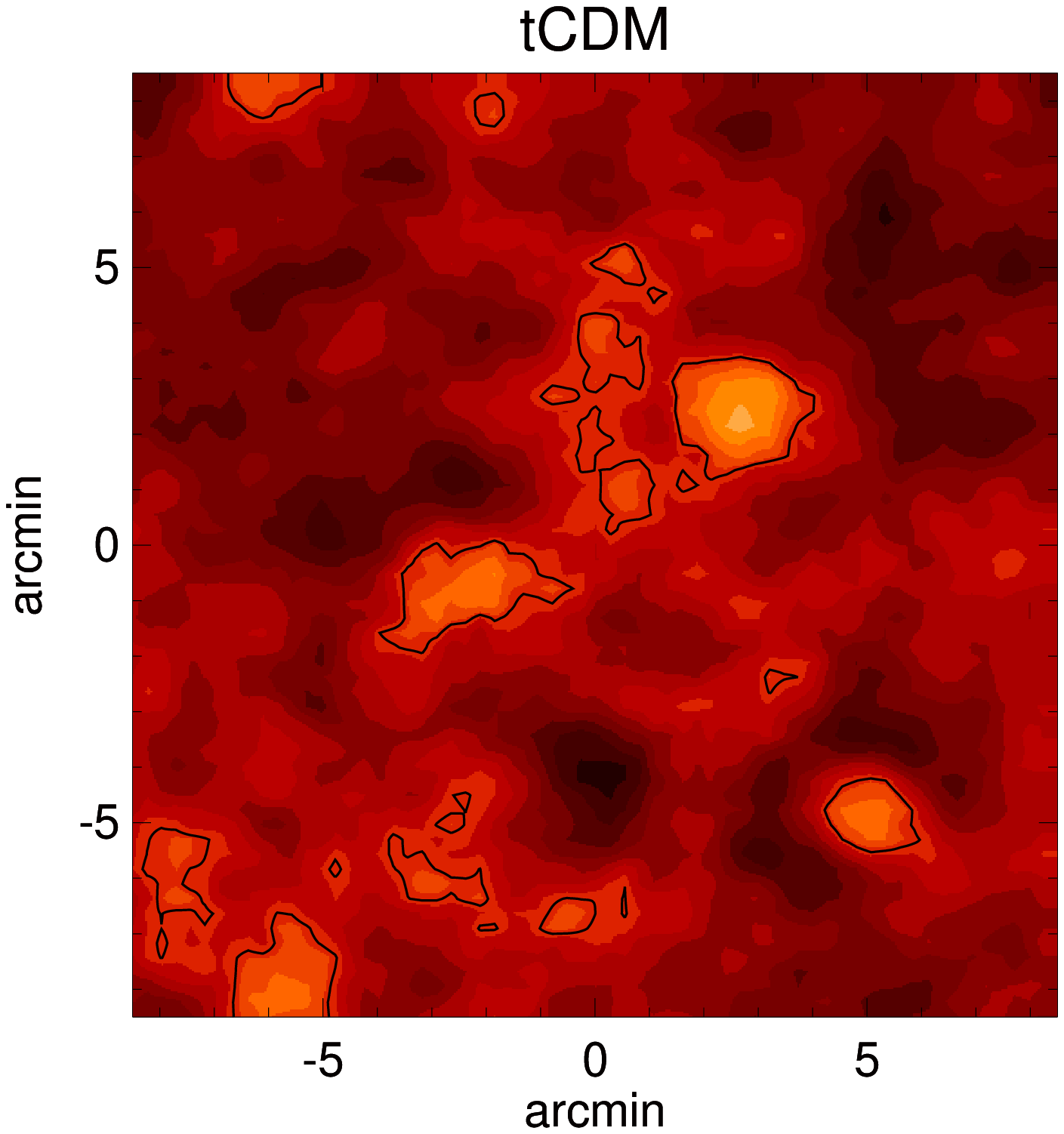,width=1.8in}
\psfig{figure=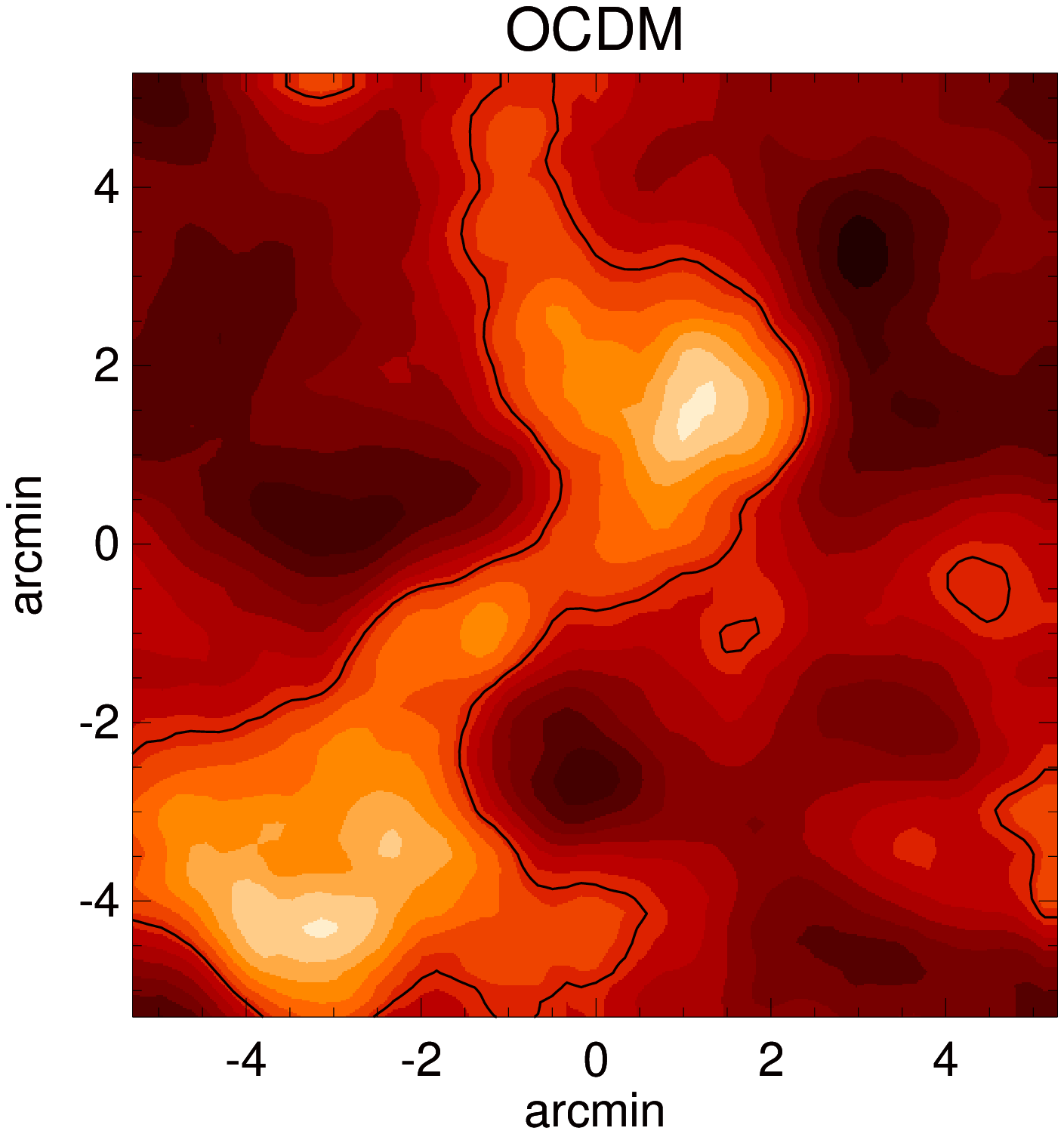,width=1.8in}}
\vspace{0.4cm}
\caption{\footnotesize {\it Left:} Radio map of redshifted 21-cm emission 
against the CMB in a `tilted' CDM (tCDM) cosmology at $z=8.5$. 
Here a collisionless $N$-body simulation with 64$^3$ particles has
been performed with Hydra (Couchman \etal 1995). 
The simulation box size is $20 h^{-1}$ comoving Mpc, corresponding to 17 (11) 
arcmin in tCDM (OCDM). The baryons are assumed to trace the dark matter
distribution without any biasing, the spin temperature to be much 
greater than the temperature of the CMB everywhere, and the gas to be 
fully neutral.
The point spread function of the synthesized telescope beam
is a spherical top-hat with a width of 2 arcmin. The
frequency window is 1 MHz around a central frequency of $150$ MHz.
The contour levels outline regions with
signal greater than $4\,\mu$Jy per beam. {\it Right:} Same for a open cosmology
(OCDM). Since the growth of density fluctuations ceases early on in an open 
universe (and the power spectrum is normalized to the abundance of 
clusters today), the signal at a given angular size is much larger in OCDM 
than in tCDM at these early epochs. (From Tozzi \etal 2000.)
}
\vspace{0.6cm}
\end{figurehere}

\ni On the theoretical side, there are several effects that need to 
be examined further. As mentioned above,
it is the presence of a sufficient flux of \Lya photons which renders the 
neutral IGM `visible'. Without heating sources, the adiabatic expansion of the
universe will lower the kinetic temperature of the gas well below that of the
CMB, and the IGM will be detectable through its absorption. If there are
sources of radiation that preheat the IGM, it may be possible to detect it 
in emission instead.
The energetic demand for heating the IGM above the CMB temperature is meager,
only $\sim 0.004\,$ eV per particle at $z\sim 10$. Consequently, even
relatively inefficient heating mechanisms may be important warming sources
well before the universe was actually reionized.
Perhaps more importantly, prior to full reionization the 
IGM will be a mixture of neutral, partially ionized,
and fully ionized structures: low--density regions will be fully ionized first, followed by regions with higher and higher densities. Radio maps at 
$21\,(1+z)\,$ cm  
will show a patchwork of emission/absorption signals from \HI zones
modulated by \HII regions where no signal is detectable against the CMB.
It is the early generation of stars 
likely responsible for reionization which will also generate 
a background radiation field of photons with energies between 10.2--13.6 eV to 
which the IGM is transparent. As each of these photons gets redshifted,
it ultimately reaches the \lya transition energy of 10.2 eV, scatters
resonantly off neutral hydrogen, and mixes the hyperfine levels. 
The observability of the pre--reionization IGM depends in this case on the 
UV spectrum of the firsts stars, i.e. on the number of \lya continuum photons 
emitted per H--ionizing photon. 

The integrated
UV spectrum of a stellar population is characterized by a strong break
at the Lyman edge whose size depends on age, initial mass function, and 
star formation
history. In Figure 4 we have therefore normalized the \lya rate to the rate 
of emission
of hydrogen--ionizing photons; after integrating over the stellar age,
$\tau$, a starburst is characterized by a ratio $N_\alpha/N_{\rm ion}\approx 
22\,(\tau/0.1\,{\rm Gyr})^{0.22}$ 
(Scalo IMF), about three times higher than the constant SFR case (this is
because of the shorter lifetime of the massive stars that produce 1 ryd
radiation).
It is possible that the startburst mode may be more relevant for the
galaxies responsible for reionization, as stellar winds and supernovae can
easily expell the gas out of the shallow potential wells of low mass halos,
after the first burst of star formation occurs. In this case the escape
fraction might be low until the gas is expelled, and as Figure 4 indicates
this would contribute to an even stronger \lya continuum  emission per 
ionizing flux.

\begin{figurehere}
\vspace{0.0cm}
\centerline{
\psfig{figure=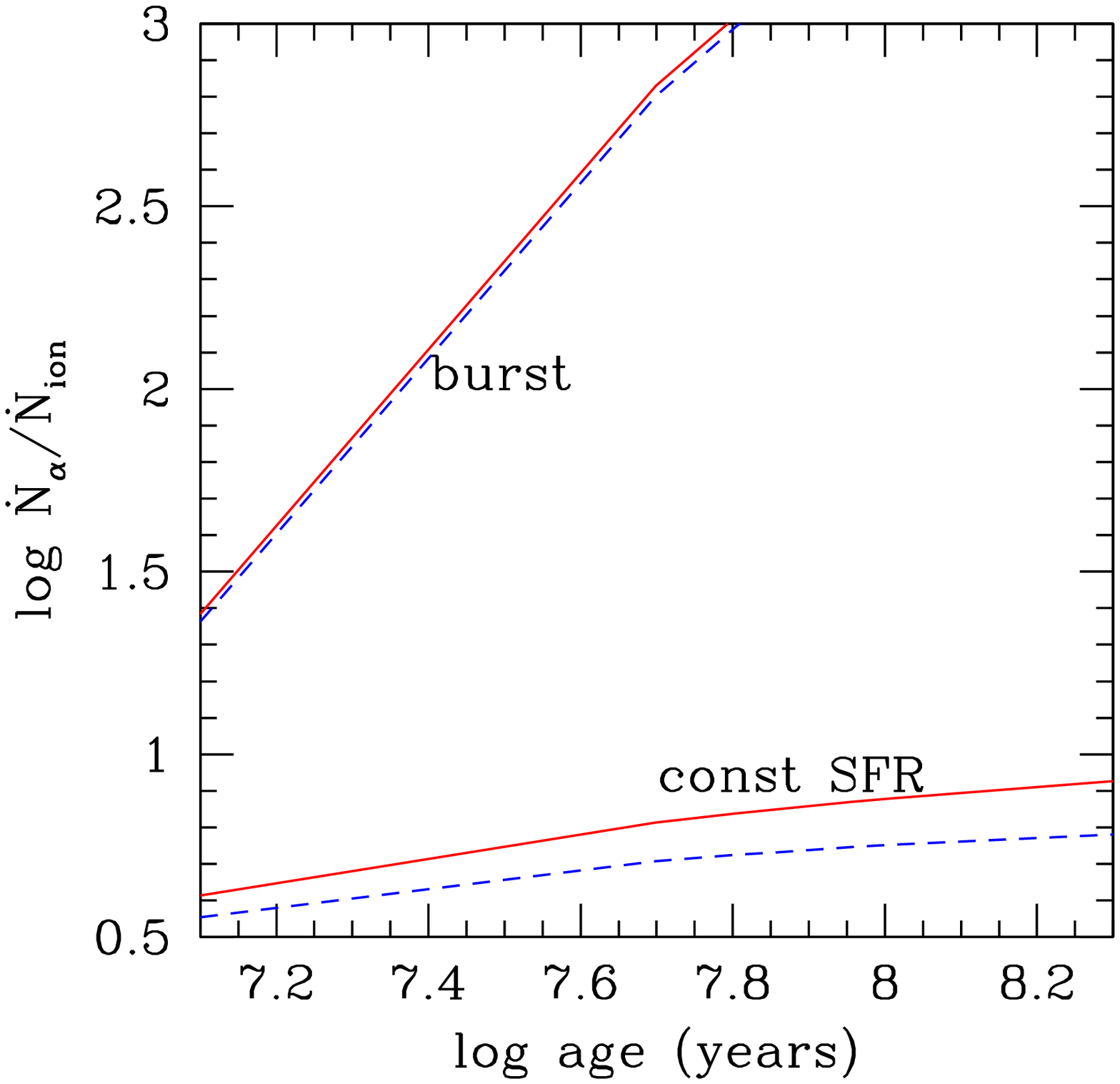,width=1.8in}
\psfig{figure=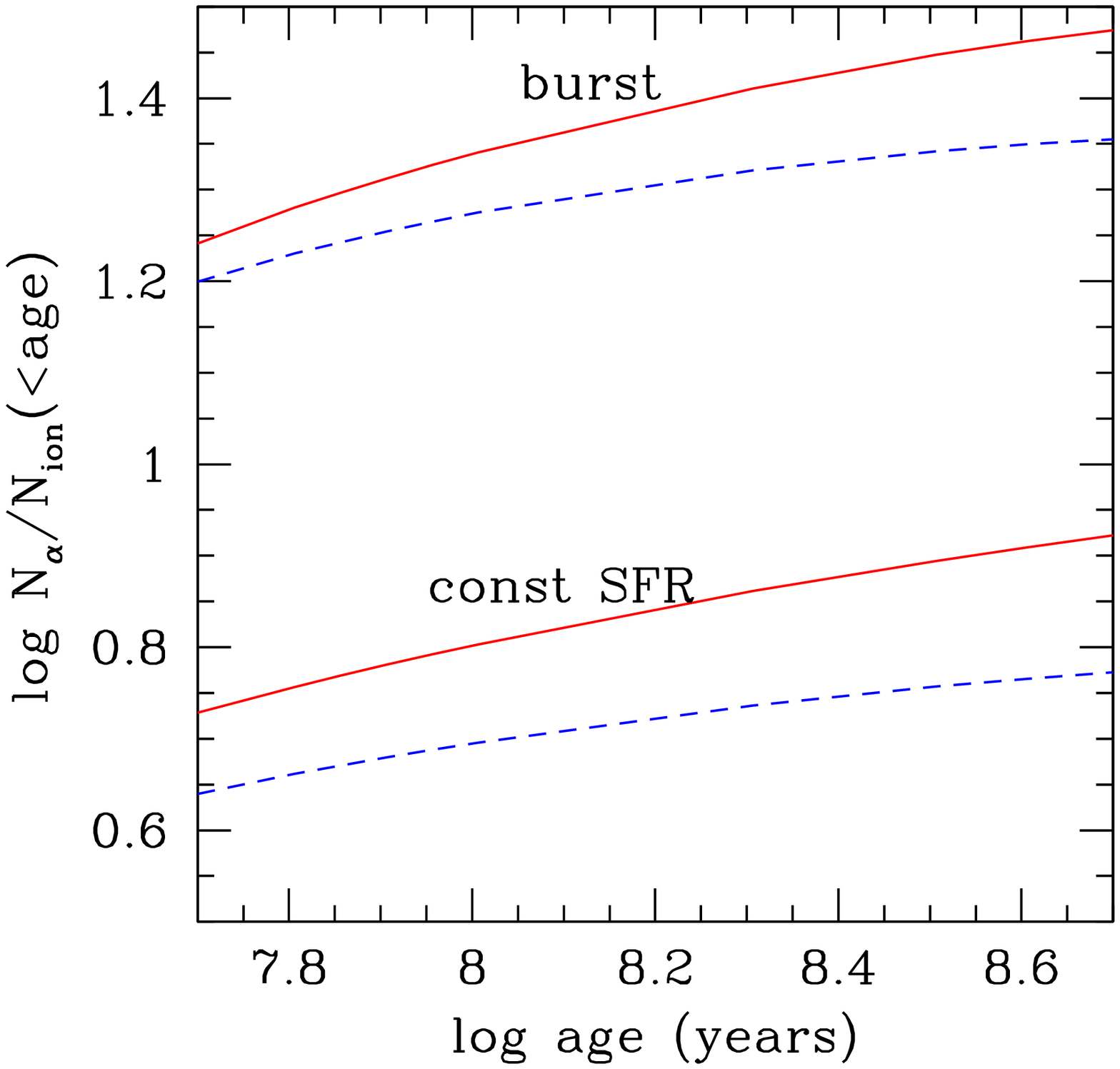,width=1.8in}}
\caption{\footnotesize 
{\it Left:} Continuum \lya photon emission rate $\dot N_\alpha$
(s$^{-1}$) versus age (in years) for a stellar population with a Salpeter
IMF ({\it dashed lines)} and a Scalo IMF ({\it solid lines}). Upper curves
assume an instantaneous burst of star formation, lower curves a constant
star formation rate. The rate of production of \lya photons has been
normalized to the rate of emission of hydrogen--ionizing photons, $\dot
N_{\rm ion}$.  The population synthesis values have been computed for low
metallicities ($Z=0.02\,Z_\odot$), and are based on an update of Bruzual \&
Charlot's (1993) libraries.  {\it Right}: Same after integration over the
age of the stellar population.
}
\vspace{0.6cm}
\end{figurehere} 

\section{Photon consumption in minihalos during cosmological reionization} 

The process of reionization has recently received much theoretical 
attention, and has been studied by several authors using semi-analytic 
models (e.g. Meiksin \& Madau 1993; Shapiro \etal 1994; Haiman \& Loeb 
1997, 1998; Benson \etal 2001) and
three-dimensional numerical simulations (Gnedin 2000).
In general, these works aim to follow the time evolution of the
filling factor of ionized (\HII) regions, based on some input
prescriptions for the emissivity and spectra of the ionizing sources.

More recent works have focused on the increased rate of recombinations
in a clumpy medium relative to a homogeneous one, i.e. when $\langle
\rho^2 \rangle > \langle \rho \rangle^2$ (Ciardi et al. 2000; Benson \etal 2001; 
Chiu \& Ostriker 2000; Gnedin 2000; Madau \etal 1999). These studies have 
left significant
uncertainties on the details of how reionization proceeds in an
inhomogeneous medium.  Since the ionizing sources are embedded
in dense regions, one might expect that these dense regions are
ionized first, before the radiation escapes to ionize the low-density
IGM.  Alternatively, most of the radiation might escape from the
local, dense regions along low column density lines of sight. In this
case, the underdense `voids' are ionized first, with the ionization of
the denser filaments and halos lagging behind (Miralda-Escud\'e et al. 2000).

At the earliest epochs of structure formation in CDM cosmologies the smallest 
nonlinear objects are the numerous small halos that condense with virial 
temperatures between the cosmological Jean temperature and $\sim 10^4\,$K. 
Such ``minihalos'' are not yet fully resolved nor is the 
process of their photoevaporation captured in large-scale three-dimensional 
cosmological simulations.\footnote{Photoionization by either 
a nearby external UV source, or by the smooth cosmic UV background radiation 
after the 
reionization epoch will heat the gas inside a minihalo to a temperature $\approx
10^4$K.  By definition, the gas inside a minihalo is then no longer
bound, leading to the photoevaporation of baryons out of their
host halos (Shapiro, Raga \& Mellema 1998; Barkana \& Loeb 1999).} 
In Haiman \etal (2001) we have used semi-analytic methods 
combined with three-dimensional numerical simulations of individual 
photoevaporating minihalos to (1) quantify the
importance of high-redshift minihalos as sinks of ionizing
photons; and (2) assess whether by extrapolating to early times the
known population of galaxies and quasars a sufficient number of UV
photons are produced for hydrogen reionization. 
If reionization occurs at sufficiently high redshifts
($z_r\gta 20$), the intergalactic medium is heated to $\sim 10^4$K
and most minihalos never form.

\begin{figurehere}
\vspace{0.5cm}
\centerline{\psfig{figure=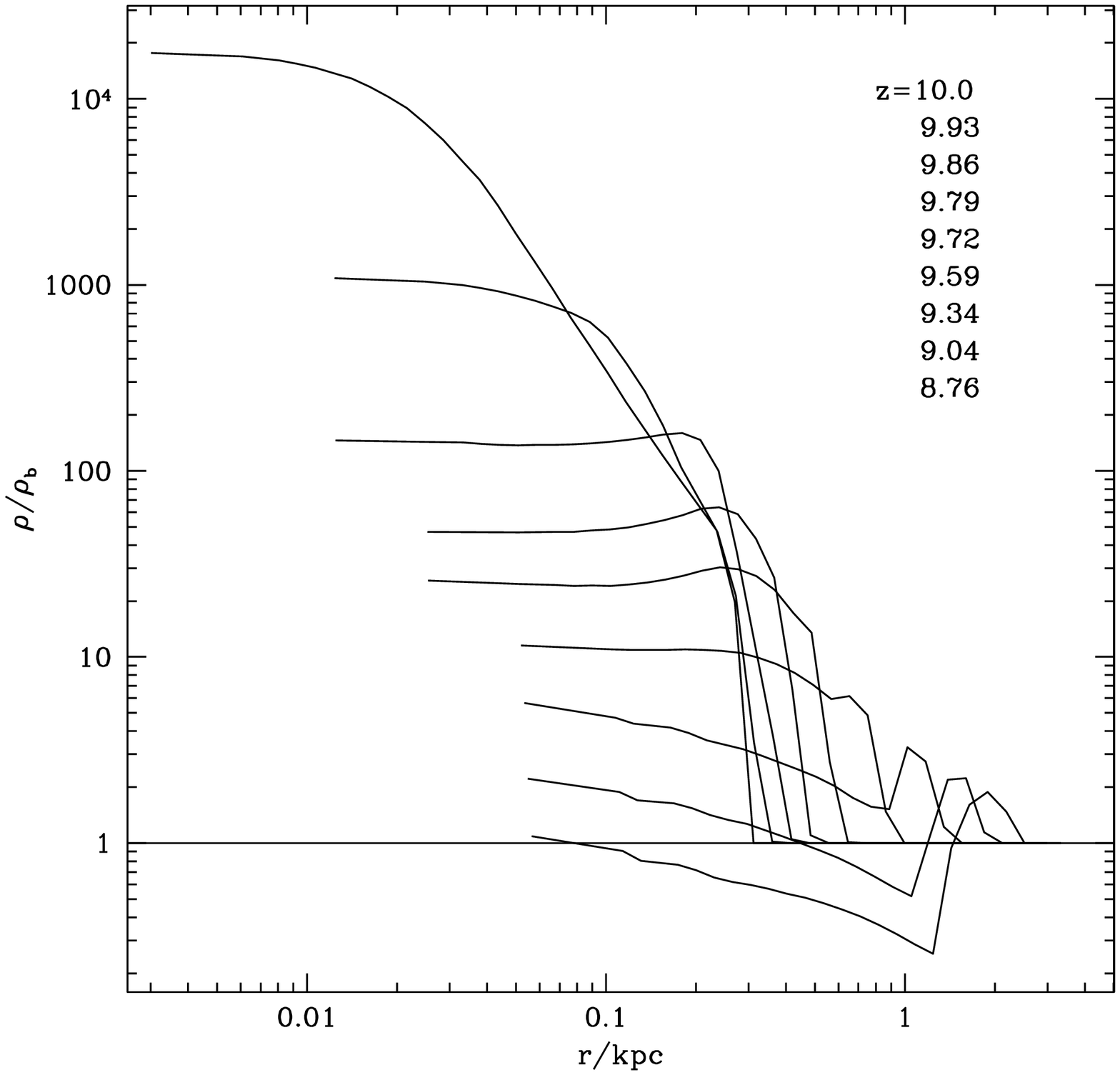,width=1.7in}
\psfig{figure=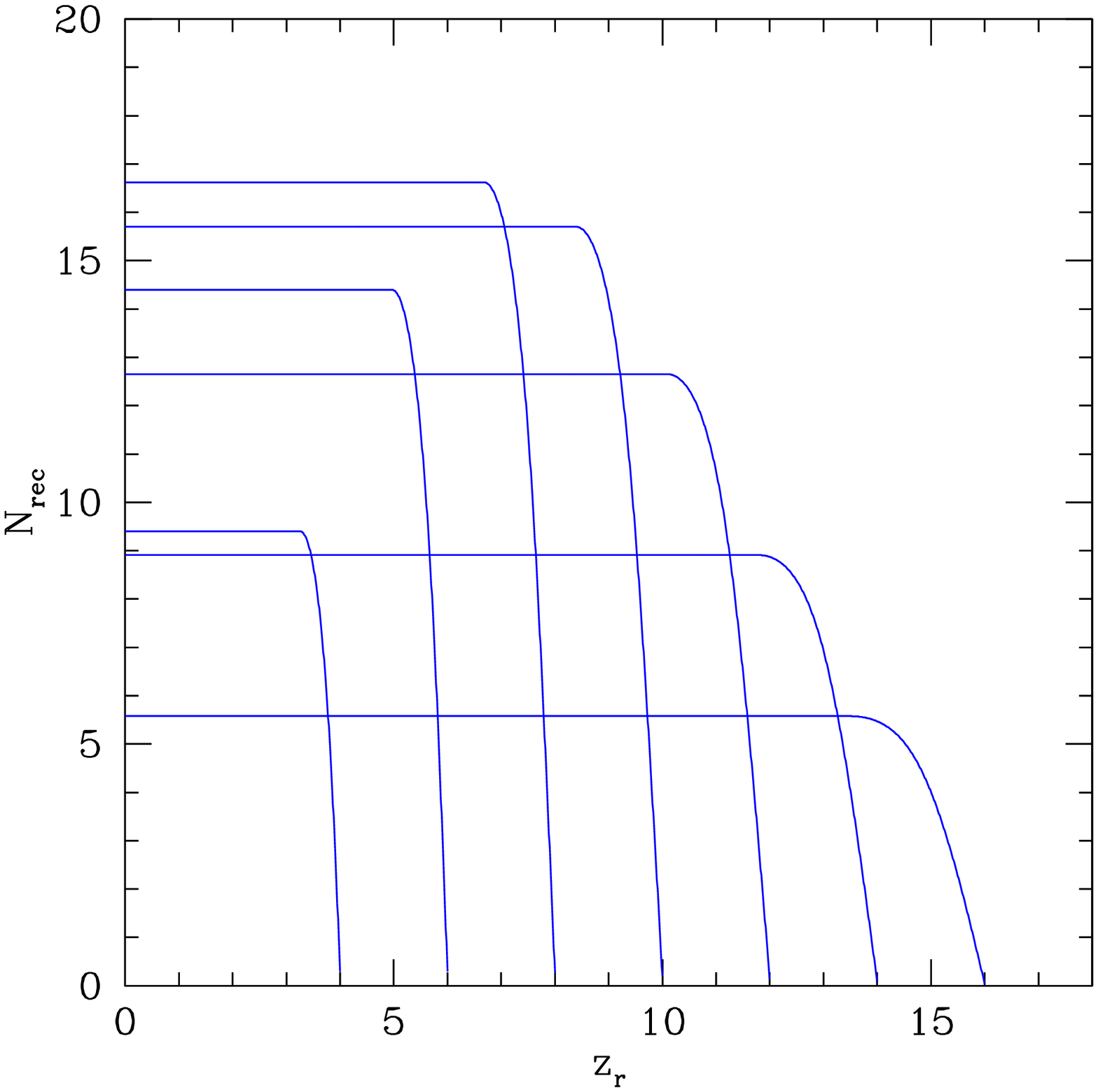,width=1.7in}} 
\vspace{0.2cm}
\caption{\footnotesize 
{\it Left:} The evolution of the
density profile of a minihalo with mass $10^6~{\rm M_\odot}$, illuminated
by a UV background flux at $z=10$. The hydrogen photoionization rate is 
assumed to be uniform on the grid, i.e. we do not solve radiative transfer.
Photoevaporation occurs very rapidly,
with most recombinations occurring within a sound-crossing time. The 
figure reveals that by $z=9.5$ the density contrast is everywhere reduced
below $\delta \lta 10$.
{\it Right:} Total number of recombinations inside minihalos, following sudden
reionization at redshift $z_r$. Recombinations are shown per background hydrogen
atom. In the earliest stages of reionization, minihalos
dominate the average gas clumping in the universe, and can therefore 
dominate the total recombination rate. 
(From Haiman \etal 2001.)
}
\vspace{0.5cm}
\end{figurehere} 

\ni On the other hand, if $z_r\lta 20$ as it now appears more likely, then a
significant fraction ($\gta10\%$) of all baryons have already
collapsed into minihalos, and are subsequently removed from the halos
by photoevaporation as the ionizing background flux builds up (Figure 5). 
This process requires a significant budget of ionizing
photons, about 10--20 of them per background hydrogen atom. This
exceeds the production by a straightforward extrapolation
back in time of known quasar and galaxy populations by a factor of
up to $\sim 10$ and $\sim 3$, respectively (Haiman \etal 2001).

\section{The earliest luminous sources and the wing of the Gunn-Peterson 
trough}

Prior to complete 
reionization at redshift $z_r$, sources of
ultraviolet radiation will be seen behind a large column of intervening gas
that is still neutral. In this case, because of scattering off the
line-of-sight due to the diffuse neutral IGM, the spectrum of a source at
$z_{\rm em}>z_r$ should show the red damping wing of the Gunn-Peterson (1965)
absorption trough at wavelengths longer
than the local \Lya resonance, $\lambda_{\rm obs}>\lambda_\alpha(1+z_{\rm em})$, where
$\lambda_\alpha=c/\nu_\alpha=1216\,$\AA.
At $z_{\rm em}\gta 6$, this characteristic feature extends for more than $1500\,\kms$
to the red of the resonance, and may significantly suppress the \lya
emission line in the spectra of the first generation of objects in the
universe. Measuring the shape of the absorption profile of the damping wing
could provide a determination of the density of the neutral IGM near the
source (Miralda-Escud\'{e} 1998).

A number of authors (Madau \& Rees 2000; Cen \& Haiman 2000; Haiman 2002) 
have recently focused on the
width of the red damping wing -- related to the expected strength of the \lya
emission line -- in the spectra of very distant QSOs and star-forming galaxies as a flag of the
observation of the IGM before reionization.
We have assessed, in particular, the impact of the photoionized, Mpc-size
regions which will surround individual luminous sources of Lyman continuum 
radiation
on the transmission of photons redward of the \lya resonance, and shown that
the damping wing of the Gunn-Peterson trough may nearly completely disappear
because of the lack of neutral hydrogen in the vicinity of a bright QSO.
The effect of this local photoionization is to greatly
reduce the scattering opacity between the redshift of the quasar
and the boundary of its \HII region (Figure 6). 

The detection of a strong \lya
emission line in the spectra of bright QSOs shining for $\gta 10^7\,$yr
cannot then be used, by itself, as a constraint on the reionization epoch. The
first signs of an object radiating prior to the transition from a neutral to
an ionized universe may be best searched for in the spectra of luminous sources
with a small escape fraction of Lyman continuum photons into the IGM, or sources
with a short duty cycle (like, e.g., gamma-ray bursts).

\begin{figurehere}
\centerline{\psfig{figure=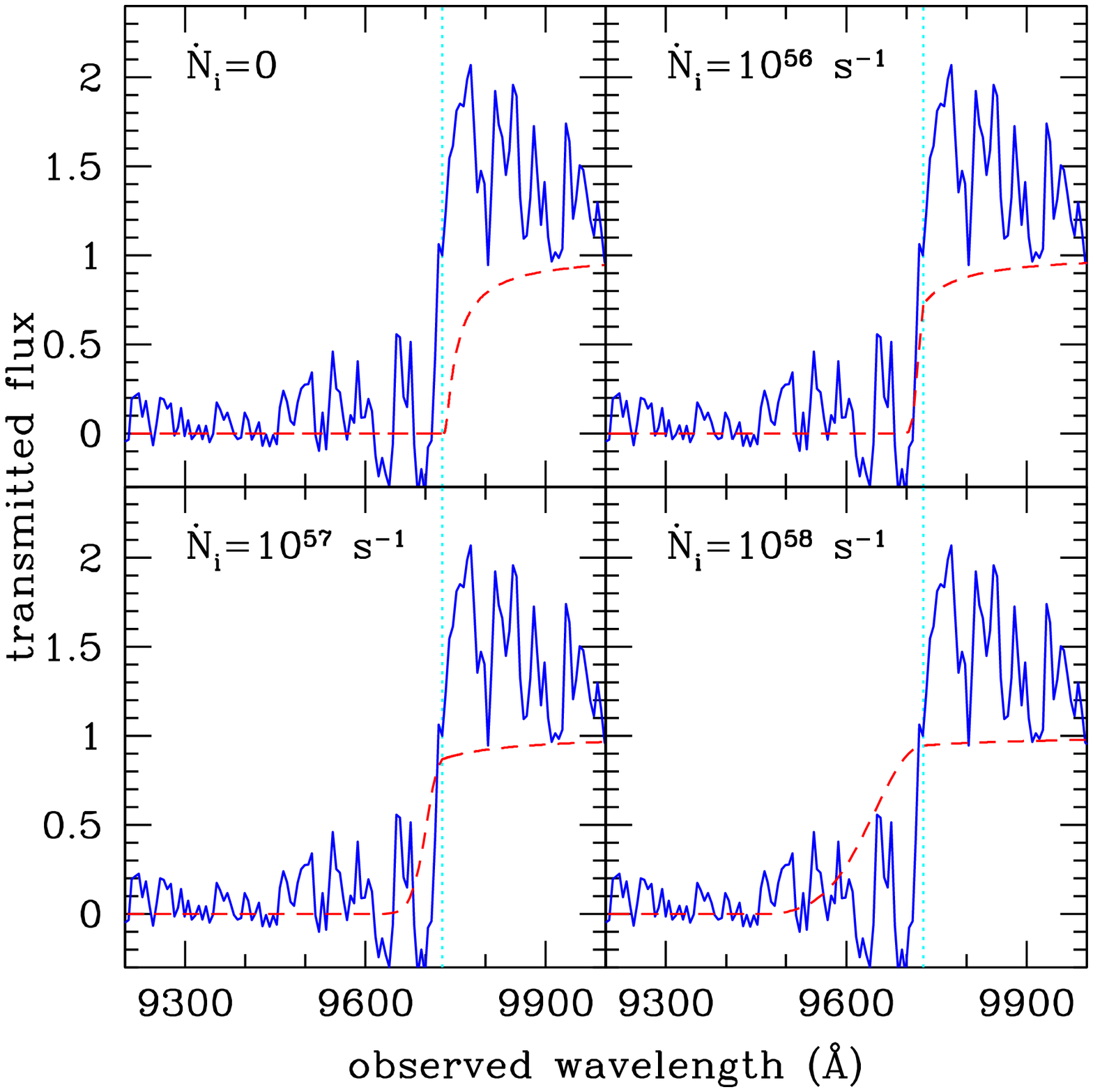,width=2.7in}} 
%\vspace{-0.9cm}
\caption{\footnotesize 
The Keck/LRIS spectrum of the faint $z_{\rm em}=5.5$ quasar RD
J0301117$+$002025 (Stern \etal 2000), redshifted to $z_{\rm em}=7$. The dotted
vertical lines indicate the expected wavelength of the \lya resonance. Fluxes
have arbitrary normalizations. The dashed curves show the transmission
$\exp(-\tau_{\rm GP})$ (where $\tau_{\rm GP}$ is the Gunn-Peterson optical
depth to resonant scattering) through a uniform IGM, assuming the QSO is being
observed prior to the reionization epoch at $z_r=6$. Four different cases 
are shown, as the rate of emission of Lyman continuum photons which ionize the 
IGM in the vicinity of the source is varied from $\dot N_i=0$ to $\dot 
N_i=10^{56}, 10^{57},$ and $10^{58}$ s$^{-1}$. The calculations assume a quasar
lifetime of $t_s=10^7\,$yr, a power-law spectrum of the form $f_\nu\propto
\nu^{-\alpha}$ with $\alpha=0.5$ near the hydrogen Lyman edge,
a recombination timescale of $t_{\rm rec}=1.3\,$Gyr, an Einstein-de Sitter 
universe with $h=0.5$, and a baryon density parameter $\Omega_Bh^2=0.02$.
(From Madau \& Rees 2000.)
}
\end{figurehere}
{}
\end{document}